\documentclass[12pt,preprint]{aastex}

\shorttitle{Discovery of a new SPB: HD~163830 from MOST}
\shortauthors{Aerts et al.}

\begin{document}

\title{Discovery of the new slowly pulsating B star HD~163830  (B5II/III) from
MOST\altaffilmark{1} spacebased photometry}

\author{C.\ Aerts\altaffilmark{2,3}, 
P.\ De Cat\altaffilmark{4},
R.\ Kuschnig\altaffilmark{5},  
J.M.\ Matthews\altaffilmark{5}, 
D.B.\ Guenther\altaffilmark{6},  
A.F.J.\ Moffat\altaffilmark{7}, 
S.M.\ Rucinski\altaffilmark{8}, 
D.\ Sasselov\altaffilmark{9}, 
G.A.H.\ Walker\altaffilmark{10}, and 
W.W.\ Weiss\altaffilmark{11}}

\altaffiltext{1}{Based on data from the MOST satellite, a Canadian Space
Agency mission, jointly operated by Dynacon Inc., the University of Toronto
Institute for Aerospace Studies and the University of British Columbia, with the
assistance of the University of Vienna.}

\altaffiltext{2}{Institute of Astronomy, University of Leuven, Celestijnenlaan
  200 B, B-3001 Leuven, Belgium, email: conny@ster.kuleuven.be}

\altaffiltext{3}{Department of Astrophysics, Radboud University
  Nijmegen, P.O. Box  9010, 6500 GL Nijmegen, the Netherlands}

\altaffiltext{4}{Royal Observatory of Belgium, Ringlaan 3, B-1180 Brussels,
  Belgium} 

\altaffiltext{5}{Department of Physics and Astronomy, University of British
Columbia, 6224 Agricultural Road, Vancouver, BC V6T 1Z1, Canada}

\altaffiltext{6}{ Department of Astronomy and Physics, St. Mary's University,
Halifax, NS B3H 3C3, Canada}

\altaffiltext{7}{D\'epartement de Physique, Universit\'e de Montr\'eal,
C.P. 6128, Succursale Centre-Ville, Montr\'eal, QC H3C 3J7, Canada}

\altaffiltext{8}{David Dunlap Observatory, University of Toronto, P.O. Box 360,
Richmond Hill, ON L4C 4Y6, Canada}

\altaffiltext{9}{Harvard-Smithsonian Center for Astrophysics, 60 Garden Street,
  Cambridge, MA 02138, USA}

\altaffiltext{10}{1234 Hewlett Place, Victoria, BC  V8S 4P7, Canada}

\altaffiltext{11}{Institut f\"ur Astronomie, Universit\"at Wien,
T\"urkenschanzstrasse 17, A-1180 Wien, Austria}

\begin{abstract}
We report the discovery of a new slowly pulsating B (SPB) star, with the largest
number of detected frequencies to date by more than a factor of three, based on
37 days of MOST (Microvariability \& Oscillations of STars) satellite guide star
photometry.  The star HD~163830 (V = 9.3, B5II/III) varies in twenty detected
frequencies in the range $0.035 - 1.06$ d$^{-1}$ ($0.4 - 12.3 ~ \mu$Hz) with
amplitudes from {0.7 to 7.6} mmag ($S/N$ from {4 to 43}).  Eighteen of these
frequencies are consistent with low-degree, high-order nonradial g-modes {of
seismic models of an evolved 4.5$M_{\odot}$ star. We are unable to identify one
unique model due to lack of mode identifications.}  The lowest two
frequencies may be associated with the rotation of HD~163830, but firm
  proof of this must await future spectroscopic data.
\end{abstract}


\section{Introduction}

Slowly pulsating B (SPB) stars (Waelkens 1991) show great promise for
asteroseismology.  Their g-mode pulsations should be sensitive probes of
their deep stellar interiors, since g-mode amplitudes remain large right
into the core (unlike p-modes whose amplitudes are significant only in
the outer layers).  But these stars also present serious observational
and theoretical challenges.  (1) Their pulsation periods are of order a
few days, making them difficult to disentangle in groundbased data due
to cycle/day aliasing.  (2) Their theoretical eigenspectra are so rich
(Dziembowski et al.\ 1993; Gautschy \& Saio 1993) that one cannot simply
find a unique match to a model by fitting the observed frequencies unless
there is prior information about the degrees and orders of the modes.
(3) Rotation is a serious complication for mode identification in these
stars because rotational splitting is large enough to cause multiplets of
adjacent radial orders to overlap.  The same great promise and obstacles
exist for other g-mode pulsators on the lower main sequence, like the
$\gamma$~Doradus variables.

Huge long-term groundbased observing campaigns (e.g., De Cat \& Aerts 2002 and
references therein for the SPBs, and Henry et al.\ 2005 and references therein
for the $\gamma$ Dor stars) had been successful in reliably discovering
multiperiodic variables. But the HIPPARCOS mission (Perryman et al.\ 1997)
demonstrated the power of spacebased observations to expand our understanding of
g-mode pulsators on the upper main sequence.  By mining the HIPPARCOS
photometric archive, Waelkens et al.\ (1998) and Handler (1999) discovered many
previously unknown SPB and $\gamma$ Dor stars, respectively. 

However, despite these efforts, the maximum number of well established
independent frequencies to be found in groundbased data of single SPB or
$\gamma$ Dor stars is five (see De Cat \& Aerts 2002; Poretti et al.\ 2002).
This paucity of observed frequencies compared to the densely populated
theoretical eigenspectra has made it difficult to even address issues like
rotational splitting.  Here, we report the detection of twenty frequencies in a
new SPB star discovered by the MOST space mission.  The WIRE satellite has
meanwhile also revealed numerous new SPB candidates with about five to fifteen 
frequencies (Br\"untt et al., in preparation).

\section{Observations and data reduction}

HD~163830 was one of 20 guide stars in the field for the MOST Primary
Science Target WR~103, observed in June 2005.  This field also contains
the guide star HD~163868 (B5Ve), in which the discovery by MOST of a
rich g-mode eigenspectrum made it the first in a new class of SPBe stars
(Walker et al.\ 2005).
The basic aspects of the guide star photometry in this field and the data
reduction procedure are already described by Walker et al.\ (2005).  A
constant star (at the level of precision of the measurements), HD~164388
(A2IV, V=7.9), was adopted as a comparison for the HD~163868 photometry,
and it is also used as a stable reference for HD~163830 in this paper.

HD~163830 was monitored for 36.6 days from 14 June to 21 July 2005, with
about 137,000 signal values in the original time series.  Data were
collected for only about 50 min of each 101-min orbit, since the target
field is outside the MOST {Continuous} Viewing Zone (CVZ) and was eclipsed
by the Earth for half of each orbit.  The final duty cycle of the time
series is about 38\%, after filtering of outliers and discarding data
contaminated by excess cosmic ray hits during satellite passages through
the South Atlantic Anomaly (SAA) in the Earth's magnetosphere.  For the
eventual frequency analysis, the data were binned into 2-min samples,
giving a total of 9890 points in the final light curve. The MOST
light curve of HD163830 is presented as dots in Fig.\,1. This plot
illustrates that the MOST data set far exceeds any previous photometric
data of known SPBs both in duty cycle and quality.

The exposure time of the star tracker was set to 1.5 sec.  As in the Direct
Imaging section of the MOST Science CCD (see Rowe et al.\ 2006), the Point
Spread Function (PSF) of the images has a Full Width Half Maximum (FWHM) of
about 2 pixels, where the focal plane scale is about 3 arcsec/pixel.  Only
pre-selected rasters of the CCD are read out, each 40x40 pixels in size, with a
guide star at the centre. The CCD parameters are: bias $\sim$550 ADU, gain
$\sim$20 electrons/ADU, and read noise $\sim$2.5 ADU/pixel RMS. A maximum of 20
such rasters can be chosen for a single guide field, due to downlink and onboard
software limitations. The brightness limits range for guide stars are from 6.5
mag to 10.5 mag.

Each raster/image is pre-processed automatically onboard. First, an
instantaneous background reading is determined by averaging the signals of the
first and last rows of the raster.  This subtracts much of the stray light
present in the image effectively, even if the scattered Earthshine changes
dramatically from exposure to exposure.  Then a fixed threshold value of 20 ADU
is added to the background; this number is subtracted from the signal from each
individual pixel. The threshold is applied to avoid introducting spurious
signals into the attitude control data processing.  As a consequence, the
contributions of the tails of the stellar PSF are not included in the
photometry.  The facts that no images are downloaded from the guide star data
and only the highest-signal pixels are included in the measurements combine to
limit the photometric precision to 2-2.5 times the photon noise limit.  However,
the long uninterrupted time series still provides low noise level results for
relatively faint guide stars such as HD~163830.

Finally, all pixels with a positive residual signal value (essentially the core
pixels of the PSF) are summed, representing a single intensity measurement per
exposure.  Residual background due to scattered Earthshine not subtracted by the
onboard procedure is filtered on the ground as described by Walker et al.\
(2005).

The MOST Startracker CCD must generally be operated at a sampling rate higher
than the Science CCD, in order to update the Attitude Control System. In the
case of the Primary Science Target WR~103, Science measurements were obtained at
10-sec intervals, during which 5 guide star exposures were accumulated.  These 5
values were added and the cumulative signal was stored in the MOST Science Data
Stream, so the effective guide star sampling interval is also 10 sec.

\section{Frequency analysis}

We performed a Fourier time series analysis using Scargle's (1982) algorithm,
successively identifying the strongest peaks and removing them from the light
curve to search for additional frequencies in the residuals.  The spectral
window of these data is so clean due to the lack of long or periodic gaps in the
time series (see Fig.~1) that aliasing is not a concern.  The prewhitening was
done by 
performing a least squares fit to find the most likely value of the 
amplitude, phase and frequency, using
the height and position of the peak 
from the Scargle periodogram as starting value. The errors
derived in this way are formal errors which do not take into account the
correlation of the noise nor the dependence among the parameters. 
Hence they are an underestimation of the true errors
(Schwarzenberg-Czerny 1991).

From this process, we identified 21 frequencies with amplitude greater
than {3.9} S/N, where the S/N ratio was estimated as 
{ the average peak amplitude in a
Scargle periodogram obtained after final prewhitening, }
in the range $0 - 2$ d$^{-1}$ where power was
detected (see Breger et al.\ 1993).   
This is considered to be {a rather strict significance criterion for the
known sample of SPBs (De Cat \& Cuypers 2003).}
One of the identified
frequencies, 2.0018 d$^{-1}$ is a known artifact in MOST data arising due to
modulation of stray Earthshine during the satellite's Sun-synchronous dusk-dawn
orbit (see Reegen et al.\ 2006). 
{ The detection of 20 independent well-separated frequencies 
within $[0,1.1]$\,d$^{-1}$ was possible thanks to the high S/N level of the
peaks and the sinusoidal waveform of the modes.}
{The satellite orbital frequency 
peak occurs at 14.2 d$^{-1}$, i.e.\
well outside the frequency range seen in the target.}

\begin{figure}
\begin{center}
\rotatebox{270}{\resizebox{12cm}{!}{\includegraphics{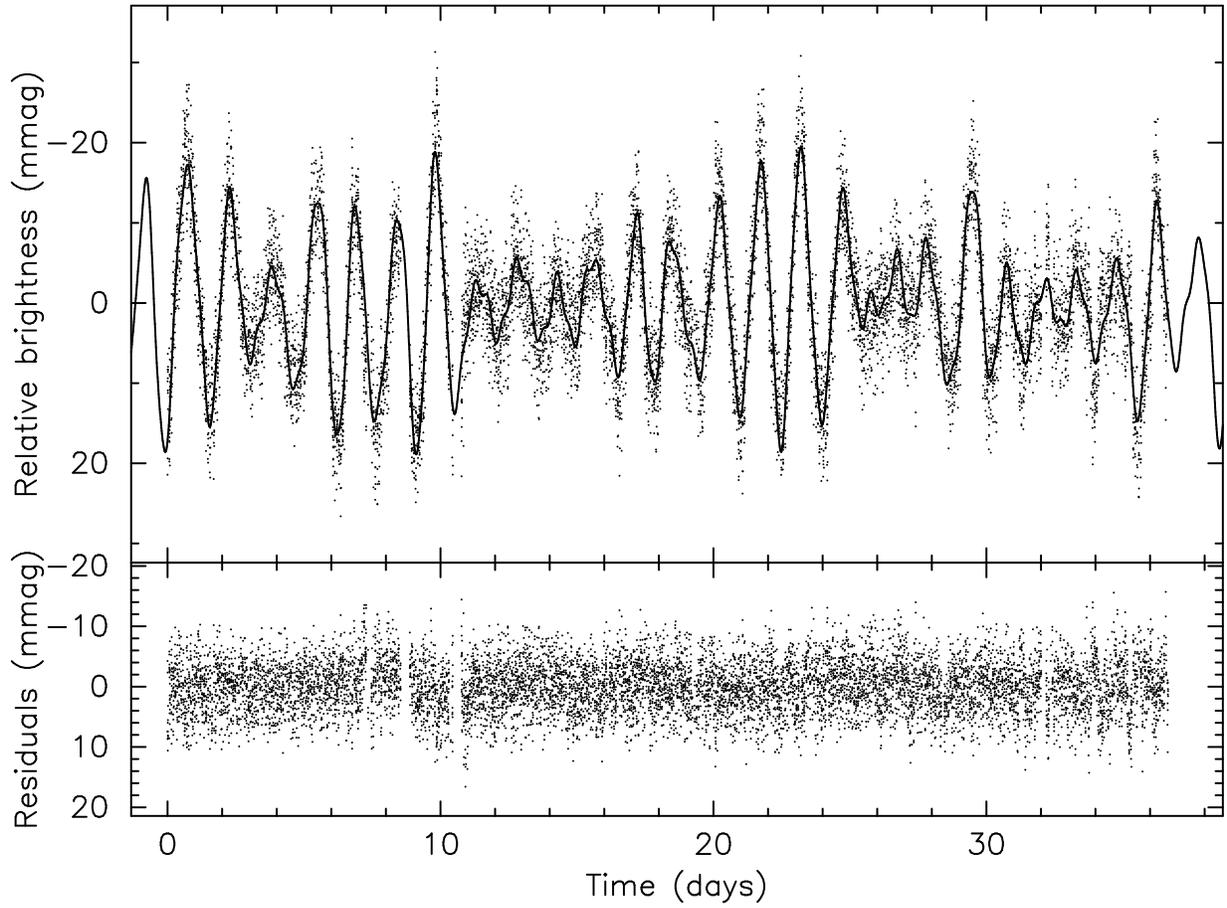}}}
\end{center}
\caption[]{Top: the observed light curve of HD~163830 (dots) is compared with
the final fit given in Eq.\,(1) (full line). Bottom: the residuals after
subtraction of the final fit.}
\label{fig1}
\end{figure}

\begin{figure}
\begin{center}
\rotatebox{270}{\resizebox{10cm}{!}{\includegraphics{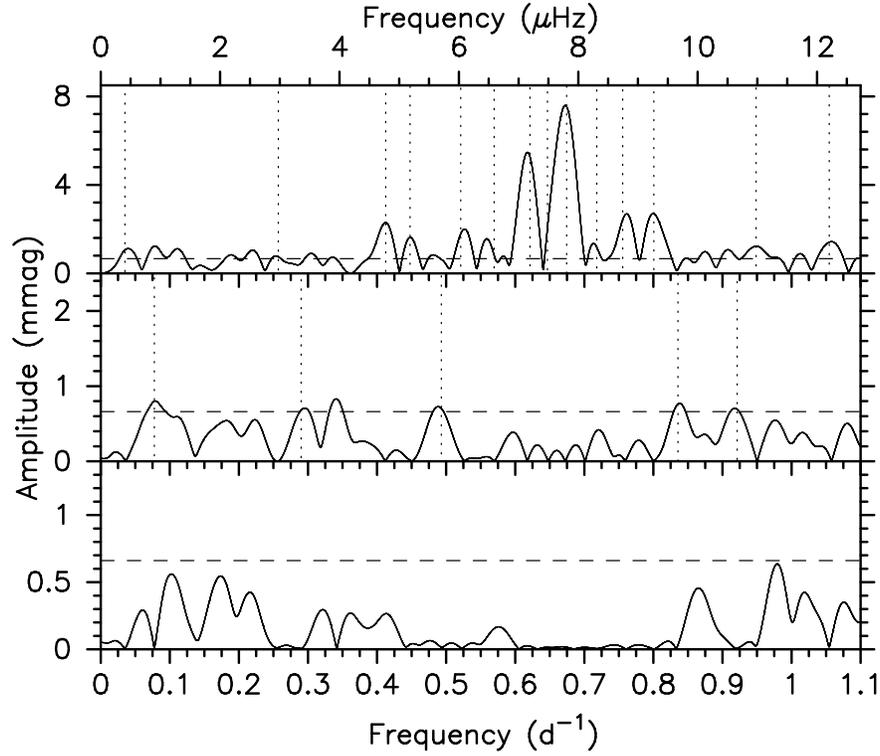}}}
\end{center}
\caption[]{Scargle periodogram of the MOST data of HD~163830 (upper panel),
  after prewhitening with the 15 dominant frequencies (middle panel) and after
  prewhitening with the 21 frequencies listed in Table\,1 (lower panel). The
  dotted vertical lines indicate the frequency values listed in Table\,1 from
  least-squares fitting. The dashed horizontal lines indicate the 3.6\,S/N
  level.
Note the different scale of the y-axes.
The stray light frequency at 2.0039 d$^{-1}$ is omitted
  for clarity.}
\label{fig2}
\end{figure}

{Scargle periodograms at a few different stages of prewhitening (omitting
the 2.0018 d$^{-1}$ artifact) are shown in Fig.~\ref{fig2}.  The final fit to
the MOST light curve of HD~163830 (including the artifact) is defined by:
\begin{equation}
y_i = a + \sum_{j=1}^{21} b_j \sin [2\pi (f_j t_i + \phi_j)],
\label{vgl}
\end{equation}
where the values for the free parameters $a, b_j, f_j, \phi_j$ are given in
Table~\ref{table1}.  The solid curve in Fig.~\ref{fig1} corresponds to this
solution.  The residuals after subtraction of this fit have a standard deviation
of 4.20 mmag and are also provided in Fig.~\ref{fig1}. Note that most of the
peaks listed in Table\,1 have an amplitude well below this value. Indeed, the
S/N ratio in the Scargle periodogram of the residuals amounts to only 0.18 mmag
mainly thanks to the reduction factor $\sqrt{9890}$.}

There is a hint for additional low-amplitude quasi-periodic
variability in the residuals. All the candidate frequencies found in them have
an amplitude below {3.6} S/N, however, so it is dangerous to make further
identifications or to assign formal errors from continued prewhitening.  We
ended the frequency identification conservatively when the formal $S/N$ dropped
below {3.9}, having high confidence in the frequencies and amplitudes in
Table~\ref{table1}.

\section{Seismic interpretation}

One high-quality archival 7-colour Geneva measurement is available for
HD~163830. We used it to derive the effective temperature and gravity of the
star following the method by K\"unzli et al.\ (1997). This resulted in $T_{\rm
eff}= 13\,700\pm 500$ K and $\log g=3.79\pm 0.14$ dex which is compatible with
the spectral type and luminosity class of B5II/III assigned by Houk (1982). As
there is no parallax estimate available for HD~163830, we interpolated in
standard models computed with the stellar evolution code CL\'ES (Scuflaire 2005)
and found a mass of 4.5$\pm$ 0.4$M_\odot$.

\begin{figure}
\begin{center}
\rotatebox{270}{\resizebox{6cm}{!}{\includegraphics{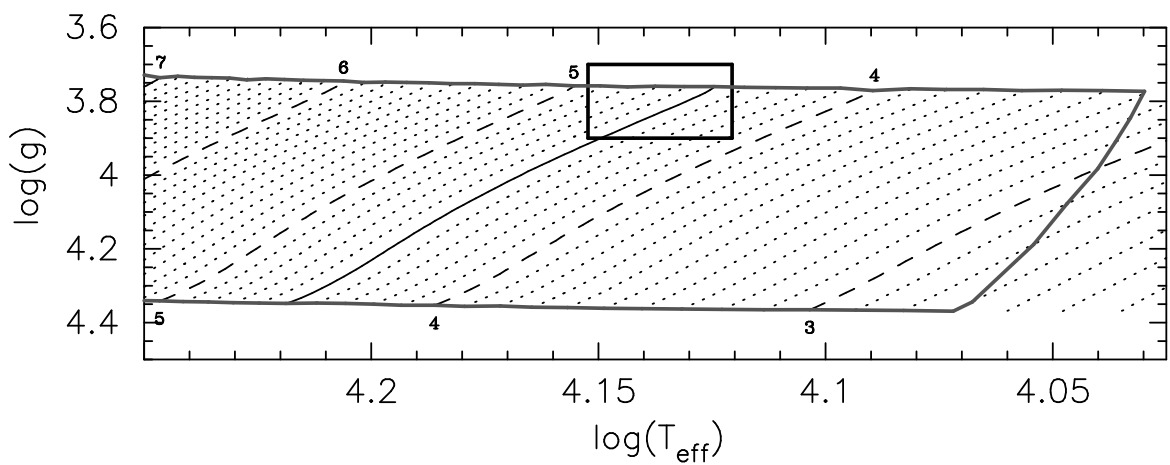}}}
\end{center}
\caption[]{Evolutionary models (dashed and dotted lines) in steps of
  0.1\,$M_\odot$ computed with CL\'ES from the ZAMS until core exhaustion. The
  masses for the models in dashed lines are given in solar values. The models
  whose frequency spectrum is given in Fig.~4 are indicated as a full black
  line. The observational error box of HD~163830 is represented by the
  rectangle. The instability strip computed from MAD is indicated by thick
  lines and represents the region in which modes of $\ell=1,2,3$ are excited. }
\label{fig3}
\end{figure}

\begin{figure}
\begin{center}
\rotatebox{270}{\resizebox{10.25cm}{!}{\includegraphics{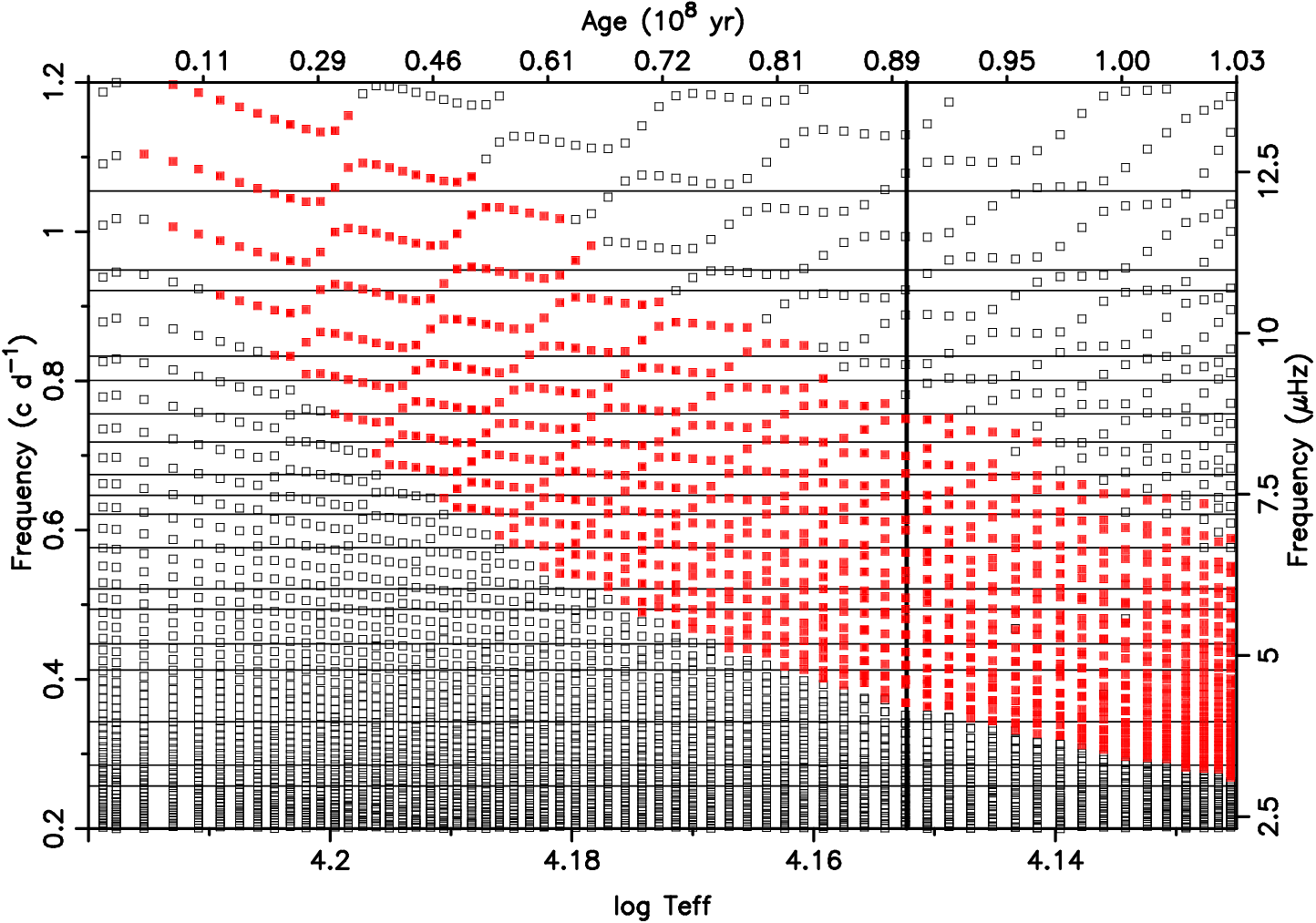}}}
\rotatebox{270}{\resizebox{10.25cm}{!}{\includegraphics{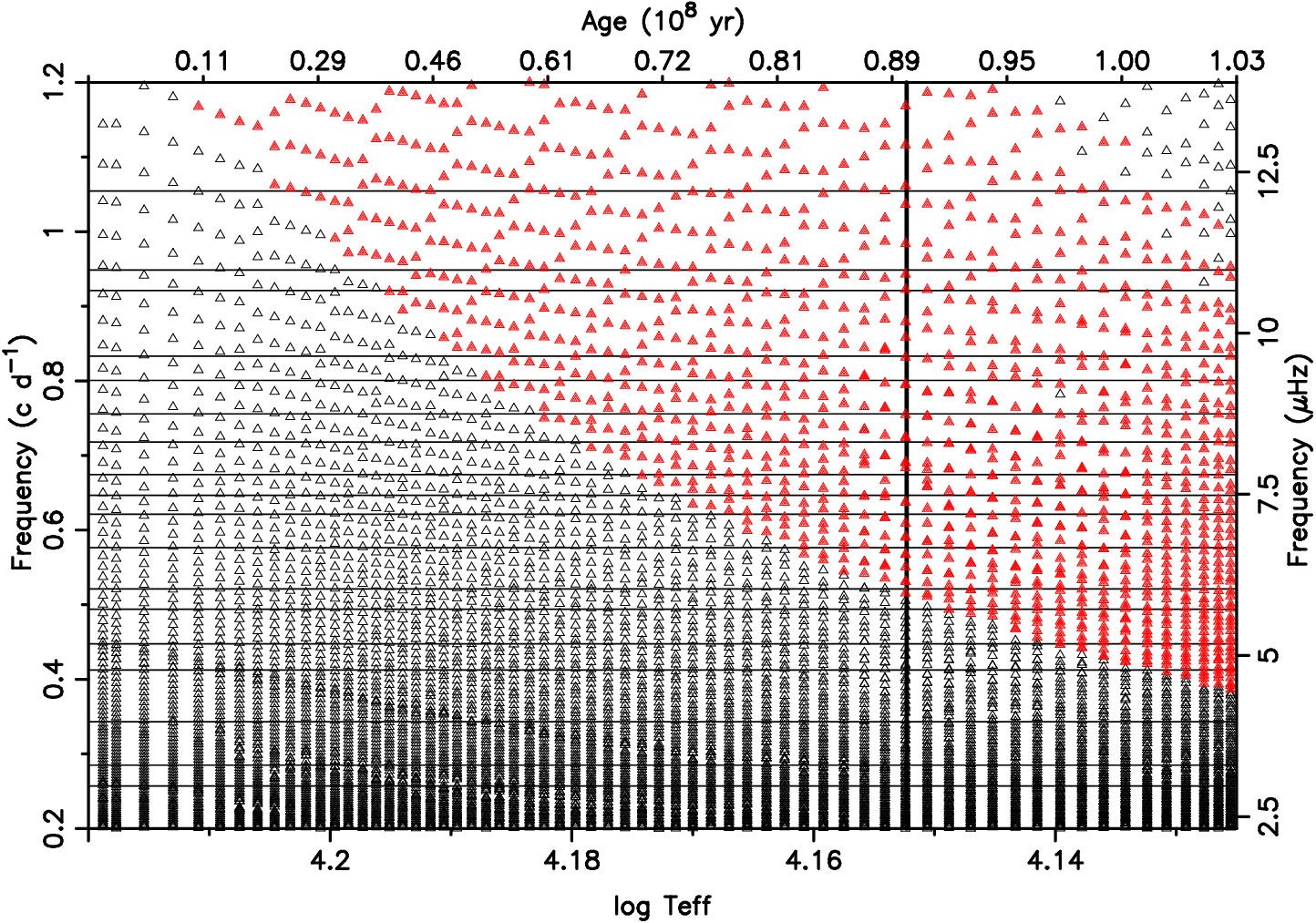}}}
\end{center}
\caption[]{Predicted frequencies for $\ell=1$ (top) and $\ell=2$ (bottom) modes
  for main-sequence stellar models with parameters 4.5$M_\odot$, $X=0.71$,
  $Z=0.015$, mixing length $\alpha=1.75$ pressure scale heights and no core
  overshooting. Excited modes are indicated with full red symbols, stable modes
  with open black symbols (colors only in the online version of the paper). The
  observed frequencies of HD~163830 are indicated as full horizontal
  lines. HD~163830 is situated to the right of the vertical line.}
\label{fig4}
\end{figure}

To compare the detected oscillation frequencies of HD~163830 with those
predicted by theory in the approximation of a non-rotating B-type star, we
computed seismic models from the ZAMS until core exhaustion using CL\'ES for 
different stellar masses and for $X=0.71$, $Z=0.015$, a mixing-length parameter of
1.75 times the pressure scale height and no core overshooting.  We plot these
models in a $(\log T_{\rm eff},\log g$) diagram in Fig~\ref{fig3}. The
observational error box of HD~163830 is also shown, as well as the SPB
instability strip for modes of $\ell=1,2,3$.  Subsequently, we computed the
frequencies of the zonal modes {of all the $M=4.5\,M_\odot$ models on the MS} for 
$\ell=1,2,3$ and tested the excitation of
these modes with the non-adiabatic code MAD (Dupret 2001). The outcome of these
instability computations is
presented in Fig.~\ref{fig4} for the $\ell=1,2$ modes. The spectrum for $\ell=3$
modes is even denser and is omitted here for brevity. We also show the observed
frequencies (except the stray light frequency and the lowest ones of {0.035} and
{0.079}~d$^{-1}$) and the upper limit of the effective temperature of
HD~163830. It can be seen that there are numerous $\ell=1,2$ modes excited in
these models and that they can explain well the observed frequencies of
HD~163830, particulary since we only show the zonal modes in Fig.~\ref{fig4}. We
thus interprete the variability of our target in terms of high-order low-degree
non-radial g-modes as expected in SPBs.  Given the expected large number of
excited modes, some of the residual power seen in Fig.\,\ref{fig1} is probably
due to g-modes we have not been able to adequately resolve from one another.
{We stress that we did not perform a matching between the observed
  frequencies and theoretical ones of the models, because we lack information on
  the $m$-values of the modes and on the stellar rotation frequency. 
Fig.~\ref{fig4} only shows that the observed spectrum is compatible with
theoretical models in the appropriate area of the HR diagram.}

The frequencies $f_{11}=0.035$~d$^{-1}$ and $f_{20}=0.079$~d$^{-1}\approx
2f_{11}$ are clearly of a different character, being an order of magnitude
smaller than the other frequencies we have identified.  A possible explanation
for these frequencies is that they are associated with the star's rotational
frequency $\Omega$. Unfortunately, the corresponding period of {28.6} days is only
slightly less than the total time coverage of the MOST photometry of HD~163830,
so it is not safe to say whether this variation even repeats periodically. Such
low frequencies have also been seen in other SPBs (De Cat \& Aerts 2002) where
they are difficult to interpret in terms of pulsation.

Some of the adjacent frequencies of HD~163830 listed in Table\,\ref{table1} show
spacings in the range $0.03 - 0.04$ d$^{-1}$, and are compatible with rotational
splitting, since high-order g-modes produce multiplets with spacings $m[1 -
1/{\ell}(\ell + 1)]\Omega$.  However, without prior mode identifications of
(${\ell},m$) it is impossible to decide whether these are multiplet components
or isolated modes.  De Cat et al.\ (2005) have shown that the oscillation
frequencies in the corotating frame of the star can differ substantially from
the same frequencies seen in the observer's frame, even in the case of slow
rotation and low-degree modes.  If $\Omega = 0.035$ d$^{-1}$, then one would
expect adjacent frequencies for $\ell = 1$ to be spaced by only 0.0175~d$^{-1}$;
for $\ell = 2$ by 0.029~d$^{-1}$.  Yet, as can be seen in Fig.\,\ref{fig4}, even
independent eigenfrequencies of adjacent radial order and the same degree in a
nonrotating model show separations comparable to this.  For these reasons, it is
premature to try to find a more specific match to {one particular} 
seismic model until there is
independent information about the rotation rate of the star or constraints on
the mode values, both of which will require spectroscopic follow up.

\section{Summary}

We discovered twenty intrinsic frequencies in the 37-day MOST light curve of the
little studied B5II/III star HD~163830. We interpreted eighteen of these
frequencies in terms of high-order non-radial g-modes in an SPB, thus adding
HD~163830 to this class of oscillators. It becomes at once the SPB with the
largest number of detected oscillation frequencies. We provided a seismic model
that is compatible with the observed frequency spectrum. This model is only one
of numerous other possibilities. In order to fine-tune our seismic analysis, we
need empirical mode identification, either from multicolour photometry or from
line-profile variations.

Our results illustrate the gain of nearly continuous photometry compared to
ground data in disentangling the frequency spectra of the g-modes in
SPBs. Similar results are to be expected for $\gamma$~Doradus stars.  We can
therefore anticipate numerous new results from space-based observatories that
are capable to gather long continuous strings of data from space, such as MOST,
WIRE, COROT and Kepler.

\acknowledgments

CA and PDC are indebted to Richard Scuflaire, Marc-Antoine Dupret and Mario
Ausseloos for providing their software codes CLES, MAD and SCAN. CA is supported
by the Research Council of the K.U.Leuven under grant GOA/2003/04. JMM, DBG,
AFJM, SR and GAHW acknowledge funding from the Natural Sciences \& Engineering
Research Council (NSERC) Canada.  RK's work is supported by the Canadian Space
Agency.  WWW received funding from the Austrian
Forschungsfo\"rderungsgesellschaft (FFG-ALR) and the Science Fonds (FWF-P17580).

\clearpage

\clearpage

\clearpage

\clearpage

\clearpage

\begin{deluxetable}{cccccc}
\tablecaption{Light curve solution of HD~163830, according to the
21 terms given in Eq.\,(\protect\ref{vgl}) besides $a=921.56(5)$ mmag.  
The reference epoch for the phases
$\phi_j$ is the time of first observation (HJD\,2453536.33498108). 
{The quoted formal errors underestimate the true variance because 
they were derived from a least-squares fit
  ignoring the correlation of the noise and the dependence among 
the parameters.}
Frequency
$f_{13}$ is due to the detection of stray light. }
\tablehead{\colhead{ } & 
\colhead{$f_j$ (d$^{-1}$)} & \colhead{$f_j$ ($\mu$Hz)} &
\colhead{$b_j$ (mmag)} & \colhead{$\phi_j$} & \colhead{S/N}} \startdata
$f_{ 1}$ &  0.6744(5) &  7.806(5)  &  7.6(2)  & 0.125(8)  & 41.4 \\ 
$f_{ 2}$ &  0.621(2)  &  7.19(2)   &  5.5(5)  & 0.34(3)   & 30.0 \\
$f_{ 3}$ &  0.647(4)  &  7.48(4)   &  2.8(4)  & 0.46(5)   & 15.3 \\ 
$f_{ 4}$ &  0.756(1)  &  8.75(1)   &  2.38(9) & 0.37(3)   & 13.0 \\
$f_{ 5}$ &  0.8007(9) &  9.27(1)   &  2.2(1)  & 0.15(2)   & 12.0 \\
$f_{ 6}$ &  0.412(1)  &  4.77(2)   &  1.64(8) & 0.50(2)   &  8.9 \\
$f_{ 7}$ &  0.949(1)  & 10.980(7)  &  1.40(6) & 0.45(2)   &  7.6 \\
$f_{ 8}$ &  0.576(2)  &  6.67(3)   &  1.4(6)  & 0.17(4)   &  7.6 \\
$f_{ 9}$ &  1.0546(7) & 12.206(9)  &  1.37(6) & 0.53(2)   &  7.5 \\
$f_{10}$ &  0.521(2)  &  6.03(2)   &  1.31(7) & 0.53(3)   &  7.1 \\
$f_{11}$ &  0.0350(8) &  0.405(9)  &  1.18(7) & 0.92(2)   &  6.4 \\
$f_{12}$ &  0.448(2)  &  5.18(3)   &  1.17(7) & 0.44(4)   &  6.4 \\
$f_{13}$ &  2.0018(8) & 23.17(1)   &  1.11(6) & 0.21(2)   &  6.0 \\
$f_{14}$ &  0.718(4)  &  8.31(5)   &  0.9(1)  & 0.61(8)   &  4.9 \\
$f_{15}$ &  0.257(2)  &  2.98(3)   &  0.9(1)  & 0.25(5)   &  4.9 \\
$f_{16}$ &  0.832(2)  &  9.63(3)   &  0.87(8) & 0.49(5)   &  4.5 \\
$f_{17}$ &  0.285(3)  &  3.30(3)   &  0.83(9) & 0.20(5)   &  4.5 \\
$f_{18}$ &  0.494(3)  &  5.72(3)   &  0.80(7) & 0.25(4)   &  4.4 \\
$f_{19}$ &  0.343(1)  &  3.97(2)   &  0.77(6) & 0.49(3)   &  4.2 \\
$f_{20}$ &  0.079(2)  &  0.91(2)   &  0.72(6) & 0.71(3)   &  3.9 \\
$f_{21}$ &  0.921(2)  & 10.66(2)   &  0.72(7) & 0.66(4)   &  3.9 \\
\enddata\label{table1}
\end{deluxetable}
\end{document}